\documentclass[aps,prx,twocolumn,amsmath,nofootinbib,amssymb,longbibliography]{revtex4-2}
\usepackage{natbib}
\usepackage{graphicx}
\usepackage{color}
\usepackage[colorlinks, allcolors=blue]{hyperref}
\usepackage{amssymb}
\usepackage{gensymb}
\usepackage{float}
\usepackage{amsmath}
\usepackage{tabularx,graphicx}
\usepackage{epstopdf}
\usepackage{latexsym}
\usepackage{color, colortbl}
\usepackage{psfrag}
\usepackage{bbm}
\usepackage{bbold}
\usepackage{bm}
\usepackage{blindtext}
\usepackage{titlesec}
\usepackage{dsfont}
\usepackage{feynmp}
\usepackage{slashed}
\usepackage{multirow}
\usepackage{appendix}
\usepackage{ragged2e}
\usepackage[normalem]{ulem}
\usepackage{coffeestains}

\newcommand{\cred}{\color{red}}

\def \ua{{\uparrow}}
\def \da{{\downarrow}}
\def \be{\begin{equation}}
\def \ee{\end{equation}}
\def \ba{\begin{array}}
\def \ea{\end{array}}
\def \bea{\begin{eqnarray}}
\def \eea{\end{eqnarray}}

\renewcommand{\vec}[1]{\boldsymbol{#1}}
\def \k {{\vec k}}

\def \bq{{\bf q}}
\def \bk{{\bf k}}

\def \e{{\epsilon}}

\def \L{{\Lambda}}

\def \t{{\theta}}

\def \g{{\gamma}}
\def \D{{\Delta}}

\def \w{{\omega}}
\def \s{{\sigma}}

\def \k{{\kappa}}

\def \e{{\epsilon}}

\def \ba{\begin{align*}}
\def \ea{\end{align*}}

\newcounter{indice}

\def \mrm{\mathrm}

\def \bs{\boldsymbol}

\date{~\today}

\begin{document}

\title{Robustness of the Kohn-Luttinger mechanism against symmetry breaking
}
\author{Amir Dalal}
\affiliation{Department of Physics, Bar-Ilan University, 52900, Ramat Gan, Israel}
\author{Vladyslav Kozii}
\affiliation{Department of Physics, Carnegie Mellon University, Pittsburgh, Pennsylvania 15213, USA}
\author{Jonathan Ruhman}
\affiliation{Department of Physics, Bar-Ilan University, 52900, Ramat Gan, Israel}

\begin{abstract}
 We investigate how strongly broken spatial symmetries affect the Kohn--Luttinger (KL) mechanism, in which superconductivity emerges purely from repulsive interactions. While the original KL argument assumes continuous rotational symmetry, real materials possess only discrete point-group symmetries, raising a central question: can sufficiently strong symmetry breaking suppress or eliminate KL superconductivity?
Using controlled perturbation theory and explicit two-dimensional models with Ising and Rashba spin--orbit coupling (SOC), we find that KL superconductivity is broadly robust and exhibits qualitatively universal behavior across models: the transition temperature $T_c$ is nonmonotonic in the symmetry-breaking field, shows a pronounced maximum at scales of the order of the Fermi energy, and decays exponentially toward zero at asymptotically large fields. 
However, the physical mechanisms determining this suppression  may differ between models. 
Overall, these results demonstrate that KL-type superconductivity can persist across a wide class of spin--orbit-coupled systems.

\end{abstract}

\maketitle
\section{Introduction}
In their seminal work, Kohn and Luttinger (KL) demonstrated that the ground state of an electronic Fermi liquid with time-reversal and translational symmetries is, in principle, superconducting~\cite{KohLuttinger}. Even when the bare interaction between electrons is purely repulsive, the static screened interaction necessarily acquires attractive components in higher angular momentum channels. This arises because the effective interaction between electrons — the scattering amplitude — contains a singularity at momentum transfer $2k_F$, where $k_F$ is the Fermi momentum. The singularity originates from Friedel oscillations, which describe the oscillatory response of a Fermi liquid to a local probe charge. The electronic density surrounding such a charge exhibits long-range ripples, and the effective potential created by the Fermi liquid inherits this spatial modulation. As a result, electron pairs with sufficiently high   orbital angular momentum can take advantage of regions in real space where the interaction becomes attractive and form Cooper pairs. While this mechanism is conceptually appealing, KL estimated in their original work that the corresponding transition temperature is vanishingly small for electrons in a typical metal. Nevertheless, the KL mechanism has recently attracted renewed attention~\cite{Erezetalrhombohedral2021,Ashvinrhombohedral2022} in the context of the experimental observation of superconductivity in rhombohedral graphene~\cite{zhou2021superconductivity,zhou2022isospin,zhang2023enhanced}.

In their analysis, KL assumed that the system possesses full rotational symmetry, such that pairing channels with different angular momenta are completely decoupled. In crystalline solids, this continuous symmetry is reduced to the discrete point-group symmetry of the lattice, leading to mixing between angular momentum channels. Only a finite number of symmetry-adapted channels remain independent. One may therefore argue that superconductivity is no longer guaranteed and may depend sensitively on the details of the bare interaction~\cite{ChubukovMaiti2013}. In the extreme limit where all spatial symmetries (except translational invariance) are broken, and all channels become strongly mixed, one might expect that  the KL mechanism will be completely suppressed. On the other hand, Friedel oscillations are much more robust — they are a generic feature of any system with a well-defined Fermi surface, regardless of rotational symmetry. These two opposing tendencies motivate the central question of this work: can sufficiently strong spatial-symmetry breaking in a Fermi liquid destroy Kohn–Luttinger superconductivity and stabilize the Fermi liquid state down to zero temperature?

In this paper,  we comprehensively address this question. We introduce a symmetry-breaking field, denoted by $\gamma$, into a range of low-energy effective two-dimensional models. The field is designed to break all point-group symmetries. Our analysis shows that the superconducting transition temperature $T_c$ remains finite for all values of $\gamma$, even when $\gamma$ exceeds the Fermi energy $\epsilon_F$. Moreover, our models typically exhibit non-monotonic behavior, with $T_c$ reaching a maximum at intermediate values of $\gamma$, which in some cases can be orders of magnitude larger than $T_c$ at the symmetric point $\gamma = 0$.
We attribute this enhancement to the impact of the symmetry-breaking field on the renormalized interaction: once rotational symmetry is lifted, the effective interaction develops a stronger momentum dependence, which in turn enhances its attractive components. 
 Eventually, as  the symmetry breaking field is taken to infinity, $T_c$ decays exponentially.   
In Ising systems, the decay originates mainly from the reduction of the density of states. In contrast, for models with Rashba SOC,  mixing between repulsive and attractive channels  plays a more central role.

The rest of this paper is organized as follows. In Section~\ref{Sec:symmetry}, we review general symmetry arguments and discuss qualitative effect of channel mixing. Section~\ref{Sec:vertex} revisits the basics of the standard second-order Kohn-Luttinger mechanism. In Sections~\ref{Sec:Isingnongeneric} and~\ref{sec:Ising}, we present specific models with Ising-type spin-orbit coupling that demonstrate resilience against crystal symmetry breaking and channel mixing. The examples with Rashba spin-orbit coupling are discussed in Section~\ref{Sec:Rashba}. We wrap up with the summary and outlook in Section~\ref{Sec:Summary}. Various technical details are delegated to the Appendices. Throughout the paper, we use units with $\hbar = k_B = 1$.

{
\section{Symmetry arguments and general expectations \label{Sec:symmetry}}
We start by reminding the basic KL symmetry argument and discussing our expectations of how spatial symmetry breaking should or should not affect this argument. As mentioned above, full scattering amplitude always has a nonanalyticity  at the  backscattering momentum transfer $q = 2 k_F$, which is a consequence of a well-defined Fermi surface. In a fully isotropic system, this scattering amplitude can be expanded in the basis of angular harmonics with the orbital angular momentum quantum number $l$. One can show that the expansion coefficients of the bare interaction, in case it is analytic and smooth, decay exponentially with $l$. The nonanalyticity of the full scattering amplitude, on the other hand, results in the power-law decay of the partial harmonics at large $l$, and at least some of them are attractive. This observation guarantees Cooper instability in some angular momentum channels, if $l$ is sufficiently large.\footnote{In many models, the strongest KL instability is already in the $l=1$ channel. In this case, however, the scattering across the whole Fermi surface contributes to the instability, and not only the vicinity of the nonanalyticity due to backscattering.} This argument, however, relies on the assumption that the system has full rotational symmetry,  ensuring that the gap equations for different angular momentum harmonics are completely decoupled.

The presence of a crystal lattice reduces continuous rotational symmetry to the discrete symmetry of the lattice’s point group. In this case, the effective interaction can still be expanded in terms of the eigenfunctions of the corresponding point group. However, only functions belonging to different irreducible representations remain decoupled in the gap equation, while the infinite set of basis functions belonging to the same representation become coupled. This coupling implies that harmonics with large angular momentum $l$ can mix with harmonics of small $l$ within the same representation. The latter may be strongly repulsive — for instance, due to the bare interaction — and can therefore completely destroy the original KL argument~\cite{ChubukovMaiti2013}. Moreover, this reasoning can be extended to the extreme limit: if all point symmetries are broken, then one may expect that even contact ($s$-wave) repulsive interaction may be sufficient to suppress the KL mechanism altogether.


To discuss these issues more concretely, we consider a generic linearized mean-field gap equation at $T = T_c$ in two dimensions 
\be\label{eq:gap_eq_intro}
\D(\t) = -\log {\L \over T_c}\int{d\phi\over 2\pi} V_{\t, \phi}\;\nu(\phi) \;\D(\phi)\,,
\ee
where the integral over the direction perpendicular to the Fermi surface has been performed.  Here 
\[\nu(\phi) = \int_0^{\infty}\frac{k dk}{2\pi}\delta(\epsilon_F-\epsilon_{\bs k}) = {\sqrt{k_F^2(\phi)+[dk_F(\phi)/d\phi]^2}\over 2\pi\,v_F(\phi)}\]
is the angular-dependent density of states per spin at the Fermi level, $\epsilon_{\bs k}$ is the single-particle dispersion, and $\epsilon_F$ is the Fermi energy.\footnote{An alternative expression for the angular-dependent density of states is $\nu(\phi) = k_F(\phi)/2\pi |\partial\epsilon_{\bs k}/\partial k|_{{\bs k} \in \text{FS}}$.} The functions $k_F(\phi)$ and $v_F(\phi) = |\partial\epsilon_{\bs k}/\partial{\bs k}|_{{\bs k} \in \text{FS}}$ are the angular-dependent Fermi momentum and velocity, respectively, and ``FS'' stands for Fermi surface. $\Lambda$ is the ultraviolet cutoff, typically given either by the Debye frequency or the Fermi energy, depending on the details. The interaction matrix is defined at the Fermi surface as $V_{\t,\phi}=V_{\bs k_F(\t), \bs k_F(\phi)}$, where $V_{\bs k, \bs p}$ is the static pairing interaction. 

When full rotational symmetry is present, $\nu(\phi) = \nu_0 = k_F/2\pi v_F$ becomes angle-independent. The interaction and the gap function in this case can be decomposed into harmonics 
\be\label{eq:fourier_decomp}
V_{\t,\phi}= \sum_{\ell} V_{\ell}\,e^{i\ell(\t-\phi)}, \qquad \D(\phi) = \sum_{\ell}\D_\ell e^{i\ell \phi}.
\ee
The gap equation~\eqref{eq:gap_eq_intro} then becomes
\be
\D_\ell = -\nu_0 V_{\ell}\log {\L \over T_c}\,\D_\ell,
\ee
which has a non-trivial solution if $V_\ell$ is attractive, $V_\ell <0$. In this case, repulsive and attractive channels of the interaction do not affect each other.

Once rotational symmetry is broken, the gap equation is modified in two essential ways. 
First, the density of states acquires an explicit angular dependence. 
Second, the interaction $V_{\theta,\phi}$, which depends on the electronic dispersion and 
the Bloch wavefunctions through screening, is no longer rotationally symmetric and thus 
cannot be written solely as a function of the angle difference $\theta-\phi$. 
As a result, different angular momentum channels become coupled in the gap equation, which 
in angular momentum space takes the form
\begin{equation}
    \Delta_{\ell}
    = - \log\!\left(\frac{\Lambda}{T_c}\right)
      \sum_{\ell_1,\ell_2}
        V_{\ell \ell_1}\,\nu_{\ell_2}\,
        \Delta_{\ell_1-\ell_2}.
\end{equation}
Such \emph{interchannel mixing} generally tends to suppress $T_c$, since it couples
attractive and repulsive components of the pairing interaction. 
However, as we will see below, anisotropy can also enhance $T_c$ by increasing the density of
states and by strengthening the  attractive eigenvalues of $V_{\t,\phi}$. 
The net effect on $T_c$ reflects the combined influence of all three ingredients 
(density of states, interaction, and channel mixing), making it difficult to draw universal
conclusions about the role of symmetry breaking in the KL mechanism without explicit calculations using concrete models.

One generic statement can nevertheless be made regarding the very existence of an attractive 
eigenvalue, and hence of a superconducting instability at sufficiently low temperature. 
Namely, if the interaction matrix $V_{\theta,\phi}$ has negative eigenvalues, then so does the 
kernel of the gap equation, $K(\theta,\phi)\equiv \nu(\phi)V_{\theta,\phi}$. 
To see this, we define
\[
\tilde V_{\theta,\phi} 
    = \sqrt{\nu(\theta)\nu(\phi)}\,V_{\theta,\phi},
\qquad
\tilde\Delta(\theta) 
    = \sqrt{\nu(\theta)}\,\Delta(\theta),
\]
so that the gap equation becomes
\[
\tilde\Delta(\theta)
    = -\log\!\left(\frac{\Lambda}{T_c}\right)
      \int \frac{d\phi}{2\pi}\, 
      \tilde V_{\theta,\phi}\,\tilde\Delta(\phi).
\]
This transformation, known as \emph{diagonal scaling}, preserves the number of positive and 
negative eigenvalues of the kernel, since it amounts to multiplication by a positive function 
$\sqrt{\nu(\theta)}>0$ (by Sylvester's law of inertia). 
Consequently, if $V$ possesses negative eigenvalues, so does $\tilde V$, ensuring that 
Eq.~\eqref{eq:gap_eq_intro} always admits a nontrivial solution. 
Of course, this argument does not fix the numerical value of the transition temperature, as our 
calculations below demonstrate.

\section{Renormalized interaction vertex \label{Sec:vertex}}
The key ingredient of Kohn-Luttinger superconductivity is the attraction in non-$s$-wave channels of the fully screened irreducible pairing interaction. Even if the bare interaction is completely repulsive, its renormalization typically leads to attraction at least in some of the channels. This effect can be most easily understood in terms of second-order perturbation theory~\cite{KohLuttinger,ChubukovMaiti2013}. 

The starting point is the bare repulsive density-density interaction 
\be  
H_I = \sum_{\bs q}U(\bs q) n_{\bs q} n_{-\bs q}, \qquad  n_{\bs q} = \sum_{\bs k, \sigma} c^{\dagger}_{\bs k + \bs q \sigma}c_{\bs k\sigma},
\label{Eq:HI}
\ee 
where $U(\bs q)>0$. Operators $c^\dagger_{\bs k \sigma}/c_{\bs k \sigma}$ are electron creation/annihilation operators and $\sigma = \uparrow,\downarrow$ is the electron spin. In the weak-coupling limit, $\nu_0 U \ll 1$, the leading corrections to the fully renormalized pairing vertex are given by the four diagrams shown in Fig.~\ref{fig:diagrams}. In the case of the contact interaction, $U(\bs q) = U_0$, and spin-degenerate Fermi surfaces, the first three diagrams cancel each other, and only the fourth ``exchange'' diagram remains. 
\begin{figure}[h!]
    \centering
\includegraphics[width=1\linewidth]{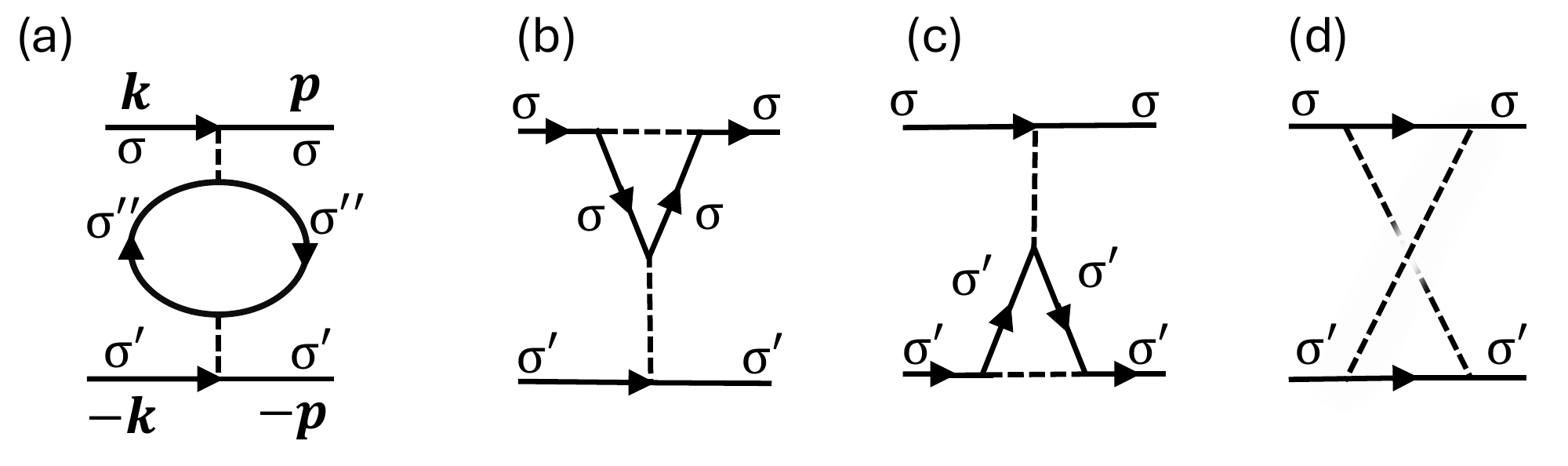}
    \caption{The four diagrams contributing to the pairing vertex at second order in the bare interaction. }
    \label{fig:diagrams}
\end{figure}
The renormalized pairing vertex in this case reads as 
\be\label{eq:V_2nd_order}
V_{\bs k, \bs p} = U_0 - U_0^2 \Pi(\bs k + \bs p)+\mathcal O(U_0^3),
\ee
where 
\be\label{eq:Pi}
\Pi(\bs q) = {T \over L^2}\sum_{i\omega,\bs k} G_0(i\omega,\bs k)G_0(i\w,\bs k+\bs q)\ee
is the static polarization bubble  and $G_0^{-1}(i\w,\bs k) = i\w-\xi_{\bs k}$ is the single particle Green's function and $\xi_{\bs k} = \epsilon_{\bs k} - \epsilon_F$. 

When $U(\bs q)$ is not constant or if the dispersion becomes spin-dependent $\epsilon_{\bs k,\uparrow} \ne \epsilon_{\bs k,\downarrow}$, the diagrams in Figs.~\ref{fig:diagrams}(a)-(c) do not perfectly cancel each other any longer and need to be calculated carefully. This procedure can be readily done~\cite{KohLuttinger,ChubukovMaiti2013}, and resulting  $V_{\bs k, \bs p}$ should be plugged into the gap equation~\eqref{eq:gap_eq_intro}, with $\bs k$ and $\bs p$ belonging to the Fermi surface. Below, we perform the calculation of the spin-dependent polarization $\Pi_{\sigma \sigma'}(\bs q)$ and analyze the gap equation for the Cooper instability in a set of two-dimensional models  with contact interaction and spin-orbit coupling. 

\section{The non-generic case: Invariance of the KL mechanism against the Fermi surfaces shift \label{Sec:Isingnongeneric}}
We begin by examining a class of models in which the Kohn-Luttinger mechanism remains completely 
unchanged, even though point-group symmetries are explicitly broken. 
These models therefore stand in stark contrast to the naive expectation that symmetry breaking 
should tend to suppress $T_c$ within the KL mechanism. 
We will show that this robustness originates from a gauge transformation that renders the 
deformed system fully equivalent to the symmetric one.

We start with a single-band spin-1/2 fermionic system having spin-degenerate Fermi surface and repulsive density-density interaction: 
\be  \label{Eq:H}
H = H_0 + H_I, \qquad H_0 = \sum_{\bs k, \sigma} \xi_{\bs k} c^\dag_{\bs k \sigma} c_{\bs k \sigma},
\ee 
and $H_I$ given by Eq.~\eqref{Eq:HI}. We assume now that this Hamiltonian has a superconducting solution through the KL mechanism. The simplest example satisfying this condition is the isotropic dispersion with the spherical Fermi surface, e.g., quadratic or quartic dispersion relations. We will discuss both of these examples in more detail below in this section.

Now consider a transformation that shifts the spin-up and spin-down Fermi surfaces at some arbitrary vectors $\bs q_\uparrow$ and $\bs q_\downarrow$, 
\be  \label{Eq:tildeH0}
\tilde H_0 = \sum_{\bs k} \xi_{\bs k - \bs q_{\uparrow}} c^\dag_{\bs k \uparrow} c_{\bs k \uparrow} + \xi_{\bs k - \bs q_{\downarrow}} c^\dag_{\bs k \downarrow} c_{\bs k \downarrow}.
\ee 
The interaction $H_I$ Eq.~\eqref{Eq:HI} is invariant under this shift. At generic vectors $\bs q_\uparrow$ and $\bs q_\downarrow$, this boost breaks all  spatial point symmetries and even the time-reversal symmetry, while the shape of each individual Fermi surface remains intact.

We can now perform a gauge transformation that compensates for these momenta shifts: 
\be  
a_{\bs k} \equiv c_{\bs k + \bs q_{\uparrow} \uparrow}, \qquad b_{\bs k} \equiv c_{\bs k + \bs q_{\downarrow} \downarrow}. 
\ee 
By simply shifting the summation variables, our Hamiltonian can now be written as 
\be  
\tilde H_0 = \sum_{\bs k} \xi_{\bs k} \left(a^\dag_{\bs k} a_{\bs k} + b^\dag_{\bs k} b_{\bs k}  \right),
\ee 
while the interaction term takes form 
\be  
H_I = \sum_{\bs q}U(\bs q) \tilde n_{\bs q} \tilde n_{-\bs q}, \,\,\,\,\,  \tilde n_{\bs q} = \sum_{\bs k} a^{\dagger}_{\bs k + \bs q}a_{\bs k} + b^{\dagger}_{\bs k + \bs q}b_{\bs k}. 
\ee 

We see that up to the change of notations $c_{\bs k \uparrow} \to a_{\bs k}$, $c_{\bs k \downarrow} \to b_{\bs k}$, Hamiltonian $\tilde H_0 + H_I$  is completely equivalent to the original one in Eq.~\eqref{Eq:H}. This means that despite all the broken symmetries in $\tilde H_0$, the system still has superconductivity through the KL mechanism, with exactly the same transition temperature. We note, however, that the actual superconducting states will have quite a different physical structure. For once, these will be  finite-momentum states (Larkin-Ovchinnikov~\cite{Larkin:1964}), with the singlet-triplet mixing on top of that. We expect that these states can potentially be distinguished from the original uniform state by, e.g., spin susceptibility or critical current anisotropy measurements. 

Now we discuss specific models with isotropic dispersion, $\epsilon_{\bs k} = \epsilon_k$, and spin-orbit coupling where this peculiar scenario may be realized. 

\subsection{Model I: Quadratic dispersion with Ising SO coupling}
The most natural case in which the scenario discussed above can be realized is the parabolic band with Ising spin-orbit coupling. We start with the spin-degenerate Fermi surface
\be  
H_0 = \sum_{\bs k, \sigma} \frac{k^2 - k_F^2}{2m} c^\dag_{\bs k \sigma} c_{\bs k \sigma},
\ee 
while the interaction term is given by Eq.~\eqref{Eq:HI} with the contact repulsion, $U(\bs q) = U_0 > 0$. 

The subtle part about this model is that it does not exhibit the KL effect in the second order of perturbation theory. The reason for this is the structure of the bare (non-interacting) static polarization operator of the fermions with parabolic dispersion in $d=2$: 
\be \label{eq:Pi2}
\Pi_2(q) = -\nu_0\!\left[1 - \mathrm{Re}\!\sqrt{1-4k_F^2/q^2 }\,\right],
\ee
with $\nu_0= m/2\pi$. Its momentum dependence is flat up to $q = 2k_F$. Since only electrons at the Fermi surface contribute to superconductivity, and the momentum transfer between them cannot exceed $2k_F$, the second-order diagrams in Fig.~\ref{fig:diagrams} simply add a small $s$-wave correction, $\propto U_0^2$, to $V_{\bs k,\bs p}$ in Eq.~\eqref{eq:V_2nd_order}. Consequently, no KL effect appears for this system in the case of a constant bare interaction at the second order. 

Nevertheless, it was demonstrated in Ref.~\cite{ChubukovKL1993} that the KL effect re-appears if higher orders of perturbation theory are considered, starting from the third one. The pairing interaction is no longer a constant at this order, with the $2k_F$ square-root non-analyticity at both sides. It results in KL superconductivity even if the bare interaction was a constant, making this model a good example to prove our point.

Next, we introduce  Ising-type spin-orbit coupling to this model, which can be viewed as planar Rashba spin-orbit where the polarization is pointing in the plane (e.g. along the $\hat y$ direction): 
\be\label{eq:H_SO_quad}
H_{SO} =  \gamma (k_x/k_0) \sigma_z,
\ee 
such that the full Hamiltonian is given by $H = H_0 + H_{SO}+ H_I$ (implying $\tilde H_0 = H_0 + H_{SO}$ in our previous notations). The effect of this term is to lift spin degeneracy and shift the Fermi surfaces according to Eq.~\eqref{Eq:tildeH0}, with $\xi_{\bs k} = (k^2 - k_F^2)/2m$ and
\be  
\bs q_\uparrow = - \bs q_\downarrow = m\gamma  /k_0\hat  {\bs x}.
\ee
where $k_0$ is a momentum scale. 
These vectors are opposite, as required by the time-reversal symmetry, see Fig.~\ref{fig:FSsandPol}(a). Parameter $k_F$ in this case remains the same, if we fix the total number of particles.

\begin{figure}
    \centering
\includegraphics[width=1\linewidth]{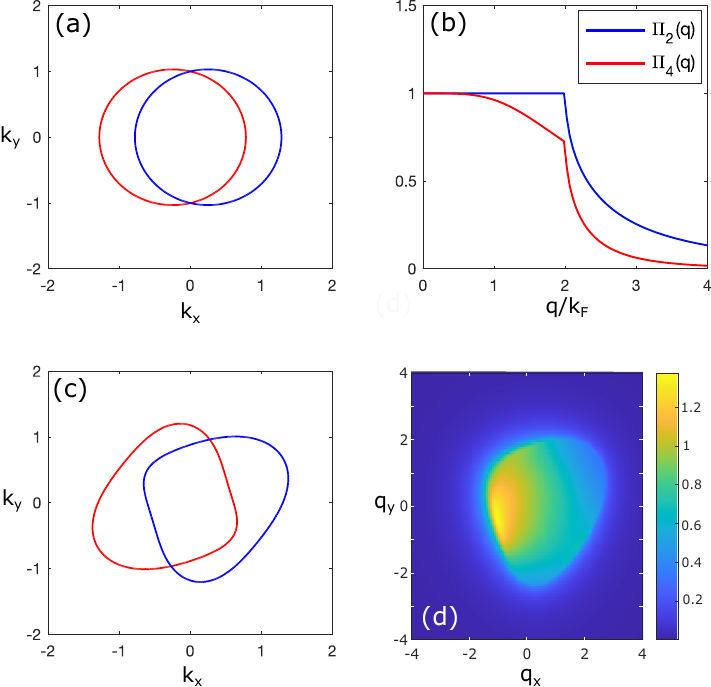}
    \caption{ (a) Fermi surfaces split by Ising spin-orbit coupling Eq.~\eqref{eq:H_SO_quad} in a quadratic system or equivalently of the Hamiltonian Eq.~\eqref{Eq:H0quartic}. The spin-orbit coupling shifts the Fermi surfaces without distorting them. (b) The polarization bubble of a parabolic and quartic dispersion relation with full rotational symmetry normalized by the density of states,  Eq.~\eqref{eq:Pi2} and Eq.~\eqref{eq:Pi4}, respectively. (c) The Fermi surfaces of the dispersion relation Eq.~\eqref{eq:xi_ising_gen} with $\gamma = \e_0$ and $\kappa = 0.4 \e_0$. (d) The polarization bubble Eq.~\eqref{eq:Pi} for the Fermi surfaces in panel (c). As can be seen it develops anisotropy and stronger momentum dependence as compared with $\Pi_4$ in panel (b).}
    \label{fig:FSsandPol}
\end{figure}

\subsection{Model II: Quartic dispersion with Ising-shifted Fermi surfaces}
To avoid the necessity of resorting to the third-order KL effect and make a connection with the rest of the paper, we now consider a model with the quartic dispersion relation $\xi_{\bs k} = \beta(k^4 - k_F^4)$:
\be  \label{Eq:H0quartic}
H_0 = \sum_{\bs k, \sigma} \beta\left(k^4 - k_F^4\right) c^\dag_{\bs k \sigma} c_{\bs k \sigma}.
\ee 
The non-interacting polarization operator of this system is now given by~\cite{Quartic2022} 
\begin{align} \label{eq:Pi4}
\Pi_4(q) = -\nu_0\frac{2 k_F^2}{q^2}
&\left[\ln\left(\frac{q^2}{2k_F^2} + \sqrt{\frac{q^4}{4k_F^4}+1}\right) \right. \\ 
- 2\Theta(q- 2k_F&)\left.\ln\left(\frac{q}{2k_F} + \sqrt{\frac{q^2}{4k_F^2}-1}\right)\right], \nonumber 
\end{align}
with $\nu_0 = 1/8\pi \beta k_F^2$ is the density of state at the Fermi level per one spin. Unlike the previous case, it has nontrivial momentum dependence even at $q<2 k_F$,  so it is sufficient to study the KL effect within the second order of the perturbation theory. 

The momentum dependence of $\Pi_4(q)$ is smooth for $0<q<2k_F$, and the $2k_F$-singularity is one-sided at $q> 2k_F$. Consequently, any conclusion about the attraction due to KL mechanism is not universal and depends strongly on the specific form of the bare repulsive interaction. Nevertheless, we can use  rotational symmetry of the system to expand effective pairing interaction on the Fermi surface in the angular momentum eigenbasis according to Eq.~\eqref{eq:fourier_decomp}: 
\be  
V_l = \int_0^{2\pi}\frac{d\phi}{2\pi} V_{\bs k_F, \bs p_F} \cos(l\phi),
\ee 
where $\phi$ now is the angle between vectors $\bs k_F$ and $\bs p_F$. Note that in this specific case the interaction vertex remains a real function when written in the band basis. Consequently, we may assume that $V_{\bs k ,\bs p} = V_{\bs p ,\bs k}$. If we assume the bare interaction is a contact, $U(\bs q) = U_0$, the pairing potential to the second order is given by Eq.~\eqref{eq:V_2nd_order}. Using that in this case $q = |\bs k_F + \bs p_F| = 2 k_F |\cos(\phi/2)|$, we obtain 
\be  
V_l = -U_0^2\int_0^{2\pi}\frac{d\phi}{2\pi} \Pi_4\left(2 k_F \left|\cos\frac{\phi}2 \right|\right)\cos(l\phi).
\ee

We find that in the case of contact interaction the leading attractive eigenvalue is in the spin-triplet $l=1$ $p$-wave channel and equals  
$V_{l=1} = - 0.074 U_0^2 \nu_0$,  with the transition temperature 
\be \label{eq:Tc_sym}
T_{c}^0 \approx  \Lambda\exp \left(  \frac1{V_{l=1} \nu_0}  \right) = \Lambda\exp \left(  -\frac1{0.074 U_0^2 \nu^2_0}  \right).
\ee 
The small numerical coefficient in $V_{l=1}$ makes it unlikely that the KL superconductivity is experimentally detectable in this specific model in the weak coupling limit $U_0 \nu_0 \lesssim 1$. However, it is well sufficient for our purposes of studying the effect of the symmetry breaking. 

For instance, we can once again break point-group symmetries simply by shifting the Fermi surfaces while keeping their circular shape intact. Time-reversal symmetry requires that the spin-up and spin-down Fermi surfaces shift in opposite directions:
\be  
\tilde H_0 = \sum_{\bs k, \sigma} \beta\left[(\bs k-\sigma \bs k_0)^4 - k_F^4\right] c^\dag_{\bs k \sigma} c_{\bs k \sigma},
\ee 
where $\sigma = +1/-1$ corresponds to spin-up/spin-down electrons, and $\bs k_0$ is an arbitrary momentum. Such a shift, which leaves the Fermi surfaces undeformed, requires fine-tuning of the model parameters but serves as a useful proof of principle. The analysis in this section shows that the transition temperature and the instabilities in all other channels remain entirely unchanged upon such shift despite the explicit breaking of spatial symmetries.

\section{The generic case: Superconductors with Ising spin-orbit coupling}\label{sec:Ising}
Models I and II discussed above, are somewhat special in that they preserve the shape of the Fermi surfaces. In this section, we turn to more generic situations where spin–orbit coupling not only shifts the centers of the circular Fermi surfaces but also distorts their shape. We start with the case where spin remains a good quantum number--Ising spin orbit coupling. 

\subsection{Gap equation for Ising superconductors}
Before turning to the calculation of $T_c$  we first briefly review some important properties of the gap equation in systems with Ising spin-orbit coupling. We start with the band Hamiltonian 
\be\label{eq:H0}
H_0 = \sum_{\bs k, \sigma} \xi_{\bs k \sigma} 
c^\dag_{\bs k \sigma} c_{\bs k \sigma}\,.
\ee
Time-reversal symmetry requires that 
\(
\xi_{\bs k \uparrow} = \xi_{-\bs k \downarrow} \equiv \xi_{\bk},
\)
so we can rewrite
\be
H_0 = \sum_{\bs k} \xi_{\bs k} 
\left(
c^\dagger_{\bs k \uparrow} c_{\bs k \uparrow}
+ 
c^\dagger_{-\bs k \downarrow} c_{-\bs k \downarrow}
\right).
\ee
Momentum \(\bs k\) runs over the full Brillouin zone.

Focusing on Cooper pairs with zero total momentum, the interaction Hamiltonian takes the form
\be
H_I = \sum_{\bs k, \bs p \in \mathrm{FS}\uparrow} 
V_{\bs k, \bs p}\,
c^\dag_{\bs k \ua} c^\dag_{-\bs k \da} 
c_{-\bs p \da} c_{\bs p \ua}.
\ee
We emphasize that the sums over \(\bs k\) and \(\bs p\) are restricted to the
spin-up Fermi surface, denoted as ``\(\mathrm{FS}\uparrow\)''.\footnote{The gap equation can be equivalently
formulated on the spin-down Fermi surface.}

Introducing the mean-field order parameter
\be \label{Eq:Deltadef}
\Delta_{\bs k}
=
- \sum_{\bs p \in \mathrm{FS}\ua} 
V_{\bs k, \bs p}\,
\langle c_{-\bs p \da} c_{\bs p \ua} \rangle,
\ee
the mean-field BCS Hamiltonian becomes
\be\label{Eq:HBCSIsing}
H_{BCS} = 
\sum_{\bs k \in \mathrm{FS}\uparrow}
\begin{pmatrix}
c^\dag_{\bs k \ua} &
c_{-\bs k \da}
\end{pmatrix}
\begin{pmatrix}
\xi_{\bs k} & -\Delta_{\bs k} \\
-\Delta^*_{\bs k} & -\xi_{\bs k}
\end{pmatrix}
\begin{pmatrix}
c_{\bs k \ua} \\
c^\dag_{-\bs k \da}
\end{pmatrix}
+ \xi_{\bs k}.
\ee
Diagonalization is straightforward via a Bogoliubov transformation; see
Appendix~\ref{app:gap_eq_Ising_SOC} for details.

The resulting self-consistent gap equation reads
\be  
\Delta_{\bs k} 
= 
- \sum_{\bs p \in \mathrm{FS}\uparrow}
V_{\bs k, \bs p}\,
\frac{\Delta_{\bs p}}{2 E_{\bs p}}\,
\tanh\!\left(\frac{E_{\bs p}}{2T}\right),
\qquad
\bs k \in \mathrm{FS}\uparrow,
\label{eq:gapeqtoytoy}
\ee
where 
\(
E_{\bs k} = \sqrt{\xi_{\bs k}^2 + |\Delta_{\bs k}|^2}
\)
is the Bogoliubov quasiparticle dispersion.

At first sight, Eq.~\eqref{eq:gapeqtoytoy} resembles the gap equation in an
inversion-symmetric superconductor.  However, it is crucial to note that the
momentum dependence of the gap is determined solely on the Fermi contour
defined by \(\xi_{\bs k\ua}=0\).  
Since this contour is generally \emph{asymmetric} and not centered at the
origin, there is no symmetry that relates the gap at antipodal points on the Fermi surface. Consequently, the condition required for the gap to be a singlet, \(\Delta_{\bs k} = \Delta_{-{\bs k}}\), does not hold, and
the pairing state is generically a mixture of singlet and triplet components.}

\subsection{Model III: Quartic dispersion with generic Ising spin-orbit coupling}
We now turn to the generic case of a shifted and distorted Fermi surface. 
To be able to study KL superconductivity within the second-order perturbation theory in the general case, we consider the model with quartic dispersion combined with a very generic Ising SO coupling: 
\be\label{eq:xi_ising_gen}
\xi_{\bs k\sigma}
= \beta k^4 
- \sigma \,\left[\gamma k_x/k_0 - \kappa k_y(3k_x^2 - k_y^2)/k_0^3\right]
- \epsilon_F,
\ee
where, again, $\sigma = +1/-1$ corresponds to spin-up/spin-down electrons. Here $\gamma$ and $\kappa$ control two types of inversion-symmetry breaking:  
$\gamma$ introduces a polar asymmetry along $k_x$, while $\kappa$ adds a threefold anisotropy that breaks all mirror symmetries.  
The Fermi energy $\epsilon_F$ fixes the electron density, defining the energy and momentum scales
$\e_0 \equiv \pi^2 \beta n^2 $ and $k_0 \equiv \sqrt{\pi n}$. 

 When $\gamma = \kappa = 0$, we reproduce Model~II given by  Eq.~\eqref{Eq:H0quartic}.
Finite $\kappa$ and $\gamma$  distort the Fermi surfaces in addition to their shifting, leading to anisotropic effective interactions. 
Figure~\ref{fig:FSsandPol}(c) shows the two Fermi contours for $\epsilon_F = \e_0$, $\gamma = \e_0$ and $\kappa = 0.4\e_0$  (blue and red correspond to the up and down spin orientations, respectively). Panel (d) shows the off-diagonal component of the corresponding polarization bubble $\Pi_{\uparrow \downarrow}$,
\be \label{Eq:Pisigmasigma'}
\Pi_{\sigma \sigma'}(\bs q) = {T \over L^2}\sum_{i\omega,\bs k} G_\sigma(i\omega,\bs k +\bs q)G_{\sigma'}(i\w,\bs k),
\ee
where $G_\sigma^{-1}(i\omega,\bs k) = i\omega-\xi_{\bs k \sigma}$ with $\sigma, \sigma' = \uparrow,\downarrow$. This is the generalization of Eq.~\eqref{eq:Pi} for the case when spin-up and spin-down Green's functions are not equivalent. Note that $\Pi_{\uparrow \downarrow}(\bq) = \Pi_{\downarrow \uparrow}(-\bq) \ne \Pi_{\downarrow \uparrow}(\bq)$ due to the presence of time reversal and broken inversion symmetry. Upon careful inspection, one can see that the polarization bubble develops a stronger momentum dependence as compared to the symmetric case,  Eq.~\eqref{eq:Pi4}. 


\begin{figure*}  
    \centering
\includegraphics[width=1\linewidth]{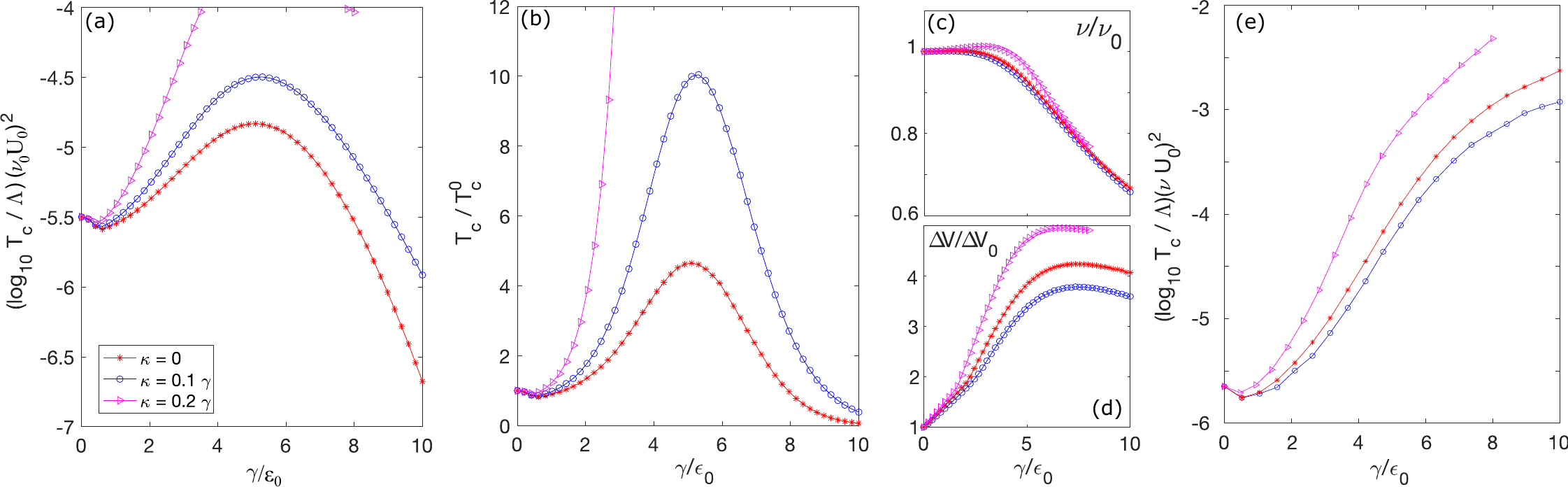}
     \caption{ Results for the KL instability for the model with Ising SOC, Eq.~\eqref{eq:xi_ising_gen}, computed with \emph{fixed density}. (a)  $\log_{10}T_c/\Lambda$ normalized by the coupling constant $(\nu_0 U_0)^2$ vs. the symmetry breaking field $\g/\e_0$ for three different values of $\kappa$ indicated in the legend and calculated for the dispersion relation in Eq.~\eqref{eq:xi_ising_gen}. Here $\nu_0$ is the density of states  per spin at the Fermi level for $\g=0$. Note that in this calculation the density of electrons $n$ is held fixed and therefore the chemical potential is changing as a function of $\g$ and $\k$. (b) The normalized $T_c/T_c^0$ vs. $\g/\e_0$. Here $T_c^0$ is the transition temperature at $\g=0$ given by Eq.~\eqref{eq:Tc_sym}. 
     (c) The density of states $\nu$ normalized by $\nu_0$ vs. $\gamma/\e_0$. (d) The difference $\Delta V$ from Eq.~\eqref{eq:DeltaV} normalized by its value at $\g=0$ vs. $\g/\e_0$.
     (e) The same as panel (a), but here we tune $U_0$ such that the product of $U_0 \nu = 0.25$ is fixed for every value of $\g$ and $\k$. This last panel shows that the origin of the decay of $T_c$ at large $\g$ is the reduction in the density of states at the Fermi level.    }\label{fig:Tc_vs_gamma}
\end{figure*}

\begin{figure}  
    \centering
\includegraphics[width=0.75\linewidth]{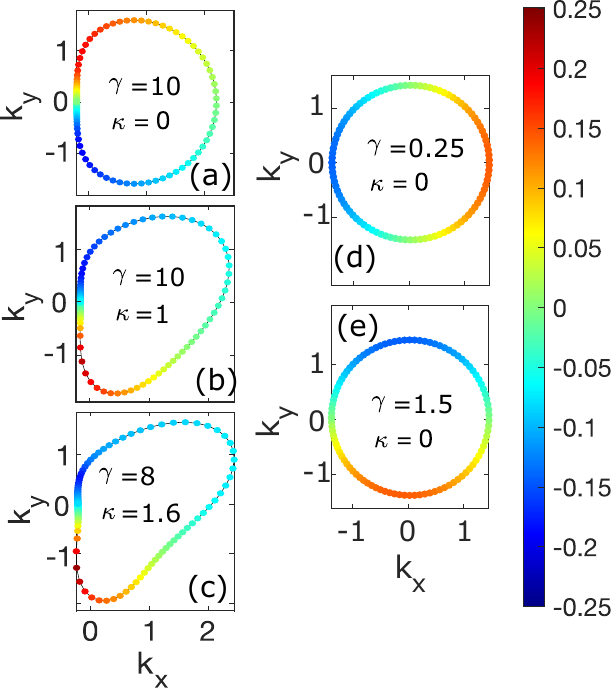}
     \caption{  The solution of the gap equation~\eqref{eq:gapeqtoytoy} overlaid on top of the spin-up Fermi surface, for various values of $\g$ and $\k$ indicated in the panels.  $\kappa$ and $\g$ are measured in units of $\e_0 = \pi^2\beta n^2 $, $k_x$ and $k_y$ are measured in units of $k_0 = \sqrt{\pi n}$, which are omitted  for brevity.  }\label{fig:FSs}
\end{figure}

In Fig.~\ref{fig:Tc_vs_gamma}, we summarize the results for the model~\eqref{eq:xi_ising_gen}, for a \emph{fixed density} of electrons and constant bare repulsion $U_0$. We have also tested the case of fixed chemical potential and the overall trends remain similar. Panel (a) displays $\log_{10}T_c/\L$, normalized by the coupling constant $(\nu_0 U_0)^2$, vs. $\g$ for three values of $\k = 0$, $0.1 \g$ and $0.2 \g$ (notice that these values depend on $\g$). Here $\nu_0$ is the Fermi level density of states per spin at the symmetric point $\g=\kappa=0$. 
In panel (b), we plot $T_c / T_c^0$ vs. $\g$, where $T_c^0$ is the transition temperature at the symmetric point given by Eq.~\eqref{eq:Tc_sym}. Panels (a)-(b) show that the finite symmetry breaking field leads to an initial decrease in $T_c$ in the regime $0<\g<\e_0$. Then, for larger values of  $\g$, we observe an increase with the maximum around $\gamma = 5\e_0$, which strongly depends on $\kappa$.  

In Fig.~\ref{fig:Tc_vs_gamma}(c), we plot the density of states
averaged over the Fermi surface and normalized to its value $\nu_0$ at the symmetric
point $\gamma = 0$, as a function of~$\gamma$. 
Panel (d) shows
\be\label{eq:DeltaV}
\Delta V \equiv 
\max_{\bs k,\bs p} V_{\bs k,\bs p}
 -
\min_{\bs k,\bs p} V_{\bs k,\bs p},
\ee
evaluated at the Fermi surface, $\bs k,\bs p \in \mrm{FS}\ua$, where $V_{\bs k, \bs p}$ is the renormalized interaction~\eqref{eq:V_2nd_order} with the off-diagonal component of the polarization operator $\Pi_{\uparrow \downarrow}$, Eq.~\eqref{Eq:Pisigmasigma'}. $\Delta V$ provides a measure for the momentum-dependent attractive component of the renormalized interaction: the gap function can exploit this momentum variation to make the sum on the right-hand side of Eq.~\eqref{eq:gapeqtoytoy} negative via an anisotropic sign structure. Since both $\nu$ and $\Delta V$ enter directly into the gap equation, we  conclude that the  $T_c$ maximum at intermediate values of $\gamma$ comes predominantly  from the combined effect of increased attraction and a reduced density of states.

As mentioned above, $T_c$ decreases slightly in the regime $\gamma \lesssim \epsilon_0$, as seen in  Figs.~\ref{fig:Tc_vs_gamma}(a)-(b). This reduction arises because of the symmetry-breaking-induced admixture of repulsive and attractive channels, which was discussed in the introduction. To see it explicitly, we resort to group-theoretical consideration and focus on the case $\kappa = 0$. This model retains only the mirror symmetry 
$k_y \to -k_y$ and therefore has the point symmetry group $C_s$. This group has 
two one-dimensional irreducible representations (irreps), $A'$  and $A''$, 
distinguished by their mirror parity: $A'$ and $A''$ are even and odd under the mirror symmetry, respectively. Importantly, the bare repulsive interaction in Eq.~\eqref{eq:V_2nd_order} is mirror-symmetric, so it occupies 
the $A'$ channel. Consequently, one may expect that the gap function in the mirror-odd irrep $A''$ is dominant, since it minimizes the overlap with the  repulsive part of the interaction.

This expectation is met at $\gamma \gtrsim \e_0$, as shown in Fig.~\ref{fig:FSs}, where we plot the numerical solution of the gap equation~\eqref{eq:gapeqtoytoy} for various $\g$ and $\k$. 
Indeed, in panel (e) we plot the solution for $\g = 1.5 \e_0$, which is clearly odd under mirror $y$ and therefore belongs to $A''$.
In contrast, at smaller $\gamma$, before $T_c$ exhibits an upturn, the leading instability is in the $A'$ irrep, which is even under $k_y\to -k_y$ and has the $p_x$ symmetry, see Fig.~\ref{fig:Tc_vs_gamma}(d). This means that  channel mixing is the subleading effect in the regime $0 < \gamma \lesssim \e_0$, while the gap structure is mainly determined by the modification of the angular-dependent density of states and the renormalized interaction due to the Fermi surface warping. Nevertheless, since the $p_x$-symmetric gap function and the bare repulsive interaction in this regime belong to the same irrep $A'$ and can mix, we do observe the decrease in $T_c$.

Finally, in the limit $\gamma \to \infty$, the transition temperature $T_c$ is exponentially suppressed in $\gamma$ for all values of $\kappa$, with the suppression becoming apparent for $\gamma \gtrsim 5\varepsilon_0$. This behavior originates from a reduction of the density of states, as shown in Fig.~\ref{fig:Tc_vs_gamma}(c). The central role of the density of states is further confirmed by rescaling the interaction strength $U_0$ by the inverse density of states, $\nu^{-1}$, computed separately for each value of $\gamma$ and $\kappa$, see Fig.~\ref{fig:Tc_vs_gamma}(e). Under this normalization, $T_c$ increases monotonically with $\gamma$. 

This observation contradicts the argument advanced in the Introduction, namely that symmetry breaking is generically detrimental to superconductivity in repulsive systems due to channel mixing. Indeed, as demonstrated in Sec.~\ref{Sec:Isingnongeneric}, Models~I and~II provide explicit counterexamples: when the density of states is held fixed, symmetry breaking leaves $T_c$ unchanged.

As we show in the next section, systems with Rashba spin-orbit coupling exhibit qualitatively similar behavior, with $T_c$ peaking at intermediate values of $\g$. However, in that case, symmetry breaking plays a more central role in suppressing $T_c$ in the $\g \to \infty$ limit.

 \section{Model IV: Quadratic dispersion with out-of-plane and planar Rashba spin-orbit coupling \label{Sec:Rashba}}
Finally, we consider systems with Rashba spin-orbit coupling where spin conservation is absent. 
\subsection{Band Hamiltonian}
We start with the band Hamiltonian of the form 
\be \label{eq:H_rashba}
H_0 = \sum_{\bs k, \sigma \sigma'} \hat{h}_{\s \s'}(\bs k )
c^\dag_{\bs k \sigma} c_{\bs k \sigma'}\,,
\ee
where
\be
\hat{h}(\bs k)=\frac{|\vec{k}|^2}{2m} + \alpha\left(k_x\sigma_y-k_y\sigma_x \right)+\gamma k_y /k_0\, \sigma_z-\epsilon_F\,,
\ee
 and in this case we define $k_0 = m\alpha$ and $\epsilon_0 = m\alpha^2/2$. 
The corresponding dispersion relation is given by 
\begin{align}\label{eq:xi_rashba}
    \xi_{\bs k,\tau}=\frac{k^2}{2m} +\tau  \alpha k\sqrt{1+r^2\sin^2\phi_{\bs k}}-\e_F,
\end{align}
where $\tau = \pm1$, $\tan \phi_{\bs k} = {k_y/k_x}$ and we have defined the ratio $r={\gamma/\alpha k_0}$.
The  matrix that diagonalizes  Hamiltonian~\eqref{eq:H_rashba} is given by

\begin{align}
\hat \L (\bs k) = \begin{pmatrix} 
\hat \L_{\ua +}(\bs k) & \hat \L_{\ua-}(\bs k) \\
\hat \L_{\da +}(\bs k) & \hat \L_{\da -}(\bs k)
\end{pmatrix},
\label{eq:FormFactors}
\end{align}
where
\begin{widetext}

\begin{align}
&\hat \L_{\ua +}(\bs k)= \cos {\beta_{\bs k}\over2}, &&\hat \L_{\ua -}(\bs k)= -\sin {\beta_{\bs k}\over2}, &&& \cos\beta_{\bs k} =\frac{r\sin{\phi_{\bs k}}}{{\sqrt{1+r^2\sin^2{\phi_{\bs k}}}}}, \nonumber \\  &\hat \L_{\da +}(\bs k)= i\sin {\beta_{\bs k}\over2} e^{i\phi_{\bs k}},&&\hat \L_{\da -}(\bs k)= -i\cos {\beta_{\bs k}\over2} e^{i\phi_{\bs k}}, &&&\sin\beta_{\bs k} =\frac{1}{{\sqrt{1+r^2\sin^2{\phi_{\bs k}}}}}.
\end{align}

In this case, there are two non-crossing Fermi contours $\bs k_F^{\tau}$. For $\e_F>0$, these two surfaces are associated with the two equations $\xi_{\bs k\tau}=0$. For $\e_F<0$ both surfaces are associated with the two non-equivalent solutions of the equation  $\xi_{\bs k-} = 0$\footnote{For $\e_F<-\e_0$ these two Fermi sheets do not enclose the origin. For simplicity, we refrain from this limit in this paper.}. As we will explain below, without loss of generality, we treat both cases as ``two bands'' denoted by the index $\tau$.

\subsection{Interaction vertex}
We now evaluate the renormalized interaction vertex.  Projecting the contact interaction
$U(\bs q) = U_0$ onto the band eigenstates at the Fermi level, we find
\begin{equation}
\label{eq:Gamma}
H_I^0 = 
\sum_{\bs k,\bs p,\bs Q}
\sum_{\tau_1,\tau_2,\tau_3,\tau_4}
\Gamma^{0}_{\tau_1\tau_2\tau_3\tau_4}(\bs k,\bs p; \bs Q)\,
\psi^{\dagger}_{\bs k+\bs Q,\tau_1}
\psi^{\dagger}_{-\bs k,\tau_2}
\psi_{-\bs p,\tau_3}
\psi_{\bs p+\bs Q,\tau_4},
\end{equation}
where the projected vertex is given by
\begin{align}
\Gamma^{0}_{\tau_1\tau_2\tau_3\tau_4}(\bs k,\bs p; \bs Q)
= \frac{U_0}{4}
\left[
\hat \Lambda^{*}_{\uparrow\tau_1}(\bs k+\bs Q)
\hat \Lambda^{*}_{\downarrow\tau_2}(-\bs k)
-
\hat\Lambda^{*}_{\downarrow\tau_1}(\bs k+\bs Q)
\hat\Lambda^{*}_{\uparrow\tau_2}(-\bs k)
\right] \nonumber \\
\times
\left[
\hat \Lambda_{\downarrow\tau_3}(-\bs p)
\hat \Lambda_{\uparrow\tau_4}(\bs p+\bs Q)
-
\hat \Lambda_{\uparrow\tau_3}(-\bs p)
\hat \Lambda_{\downarrow\tau_4}(\bs p+\bs Q)
\right].
\end{align} 
The renormalized vertex at the lowest non-trivial order is represented by the four diagrams shown in Fig.~\ref{fig:diagrams}. As in the previous section, diagrams (a), (b), and (c) cancel each other in the case of contact (momentum-independent) bare interaction, so only diagram (d) remains. In the Cooper channel with zero center-of-mass momentum, it takes the following form:
\begin{equation}\label{eq:Gamma}
\Gamma_{\tau\tau'}(\bs k,\bs p) =
\Gamma^{0}_{\tau\tau\tau'\tau'}(\bs k,\bs p;0)
-
\sum_{ q, \tau_1,\tau_2}
\Gamma^{0}_{\tau'\tau_2\tau\tau_1}(\bs p,\bs p+\bs k-\bs q;\bs q-\bs p)
\Gamma^{0}_{\tau_1\tau'\tau_2\tau}(\bs p+\bs k-\bs q,\bs k;\bs q-\bs k)
G_{\tau_1}(q_0,\bs q)G_{\tau_2}(q_0,\bs q-\bs p-\bs k),
\end{equation}
 with $G^{-1}_{\tau}( k) = ik_0 - \xi_{\bs k,\tau}$ and the sum over $q=(q_0,\bs q)$ includes a summation over the fermionic Matsubara frequencies $q_0$.
\begin{figure*}
    \centering
    \includegraphics[width=1\linewidth]{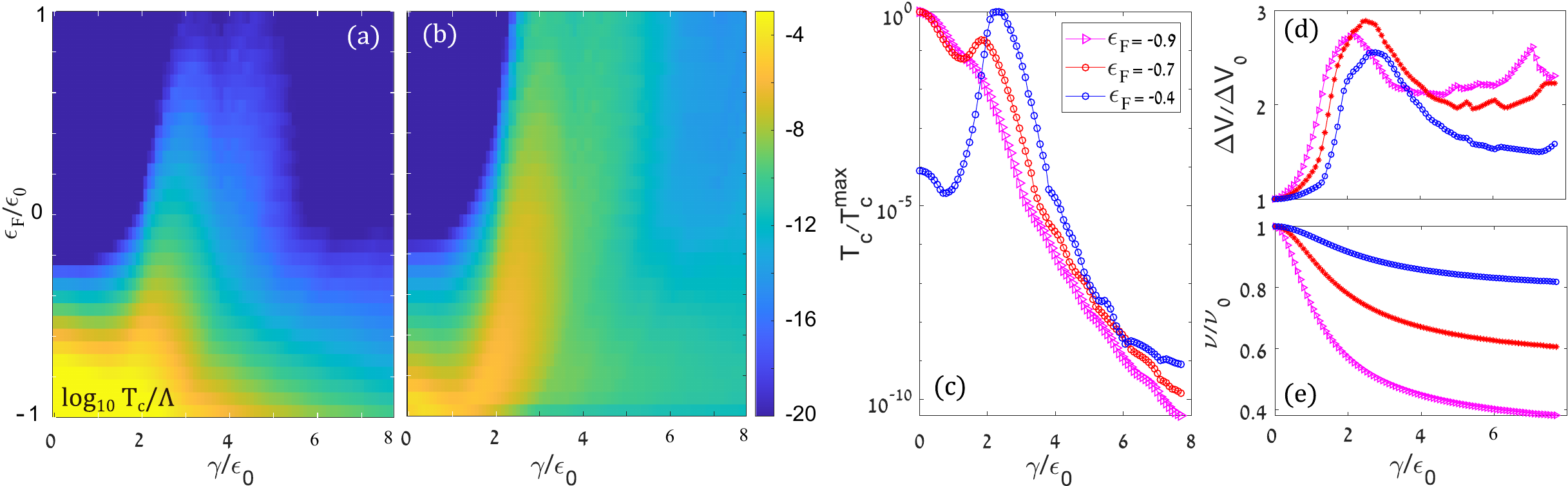}
    \caption{Results for the model with Rashba spin–orbit coupling, Eq.~\eqref{eq:H_rashba}, computed at \emph{fixed chemical potential}. 
(a) Heat map of $\log_{10}(T_c/\Lambda)$, obtained from Eq.~\eqref{eq:gap_eq_rashba}, shown as a function of the Fermi energy $\epsilon_F/\epsilon_0$ and the symmetry–breaking field $\gamma/\epsilon_0$, for $U_0 m/2\pi = 0.15$. 
(b) The same quantity, $\log_{10}(T_c/\Lambda)$, computed instead at fixed $U_0 \nu = 0.15$, where $\nu$ is the density of states at Fermi level for a given $\g$ and $\e_F$. This panel therefore isolates the influence of $\Delta V$ and channel mixing by removing variations due to changes in the density of states.
    (c) $T_c / T_c^{max}$ for three slices of panel (a), $\e_F/\e_0 = -0.9,\,-0.7$ and $-0.4$, as a function of $\g/\e_0$. $T_c^{max}$ is the maximum value of $T_c$ in the plotted range.  (d) The density of states averaged over the Fermi surfaces normalized by $\nu_0$ as a function of $\g/\e_0$. $\nu_0$ is the average density of states per spin at the symmetric point $\g = 0$. (e)~The measure of attraction strength $\D V / \D V_0$ vs. $\g / \e_0$.}
    \label{fig:Tc}
\end{figure*}
The antisymmetrized pairing vertex is then given by 
\begin{equation}
V_{\tau\tau'}(\bs k,\bs p) =
\frac{1}{4}\left[
\Gamma_{\tau\tau'}(\bs k,\bs p)
-\Gamma_{\tau\tau'}(-\bs k,\bs p)
-\Gamma_{\tau\tau'}(\bs k,-\bs p)
+\Gamma_{\tau\tau'}(-\bs k,-\bs p)
\right].
\end{equation}
Because we only consider Cooper pairs with zero total momentum, the pairing is purely intraband, so the pairing vertex must satisfy
$V_{\tau\tau'}(\bs k,\bs p)
=-V_{\tau\tau'}(-\bs k,\bs p)
=-V_{\tau\tau'}(\bs k,-\bs p)$. Note that this assumption is only valid as long as the symmetry breaking field is significantly larger than the zero temperature pairing gap, i.e., $\D\ll \g$.
When $\epsilon_F < 0$, both Fermi sheets belong to the $\tau=-$ band; in this case we simply use
$\tau=\pm$ to label the outer and inner sheets of the same band, with the corresponding form factors
$\Lambda_{\sigma,-}(\bs k_F^\tau)$.
When $\epsilon_F > 0$, the two Fermi sheets originate from different bands and the labels
$\tau=\pm$ coincide with the physical band indices.  
In both regimes, we therefore use $\tau=\pm$ uniformly in the expression above to index the two Fermi surfaces ($\tau = -$ for the larger Fermi surface and $\tau= +$ for the smaller Fermi surface).

\subsection{Gap equation and calculation of $T_c$}
The gap equation, Eq.~\eqref{eq:gap_eq_intro}, now takes the form  
\begin{equation}
\label{eq:gap_eq_rashba}
\Delta_\tau(\theta) 
= -\log \frac{\Lambda}{T_c}
\sum_{\tau'}\int_0 ^{2\pi} \frac{d\phi}{2\pi}\,
V_{\tau \tau'}(\bs k_F^{\tau}(\t),\bs k_F^{\tau'}(\phi))\, 
\nu_{\tau'}(\phi)\,
\Delta_{\tau'}(\phi)\,.
\end{equation}
\end{widetext}
The antisymmetrization of $V_{\tau,\tau'}(\bs k,\bs p)$ above ensures that the gap function is also antisymmetric, i.e., $\D_\tau(\t+\pi ) = -\D_\tau(\t)$, in agreement with Ref.~\cite{gor2001superconducting}.

Figure~\ref{fig:Tc}(a) shows $\log_{10}(T_c/\Lambda)$, computed from Eq.~\eqref{eq:gap_eq_rashba}, as a function of the Fermi energy $\epsilon_F/\epsilon_0$ and the symmetry-breaking field $\gamma/\epsilon_0$ for $mU_0/2\pi = 0.15$. In contrast to Fig.~\ref{fig:Tc_vs_gamma}, the chemical potential is held fixed, so the density varies with~$\gamma$. The overall trend resembles that of Fig.~\ref{fig:Tc_vs_gamma} in Section~\ref{sec:Ising}: following a shallow dip, $T_c$ increases with $\gamma$, reaches an intermediate maximum, and then decreases exponentially. A closer inspection, however, reveals several important differences.

Figures~\ref{fig:Tc}(b)–(e) clarify the comparison with Section~\ref{sec:Ising}.  
Panel~(b) again shows $\log_{10}(T_c/\Lambda)$ in the $(\gamma,\epsilon_F)$ plane, but with $U_0\nu$ fixed to $0.15$, where $\nu$ is the density of states per spin at the Fermi level computed for each value of $\g$ and $\e_F$. This removes variations arising from the changing density of states and highlights the role of channel mixing and variations in the renormalized interaction. Viewed together with panel~(d), which shows that $\Delta V$ in Eq.~\eqref{eq:DeltaV} does not decay as $\gamma \to \infty$, the suppression of $T_c$ due to channel mixing becomes evident.

In panel~(c), we plot three slices of the heat map in panel~(a), $\epsilon_F/\epsilon_0 = -0.9,\,-0.7,\,-0.4$, normalized by their peak values $T_c^{\text{max}}$, as a function of $\gamma$. For $\epsilon_F = -0.4\epsilon_0$, the behavior mirrors that of Fig.~\ref{fig:Tc_vs_gamma}: $T_c$ peaks at intermediate $\gamma$, reflecting competition between the enhanced attractive contribution $\Delta V$ [panel~(d)] and the reduced density of states [panel~(e)]. At very low densities $\e_F = -0.7\epsilon_0$ and $-0.9\e_0$, however — where $T_c$ is largest — the trend differs: the global maximum occurs at $\gamma = 0$, although a secondary local maximum may appear at intermediate $\gamma$. The maximum at the symmetric point, $\gamma=0$, results from the strong dependence of the density of states on $\g$ at small $\e_F$, as shown in panel (e).\footnote{For $\epsilon_F < 0$, the density of states diverges as $\nu = m/\pi\sqrt{1+\e_F/\e_0}$ at $\e_F \to -\e_0$ and the polarization bubble exhibits nontrivial momentum dependence for $0 < q < 2k_F^-$ even when $\gamma = 0$~\cite{ruhman2014ferromagnetic}.}

Finally, for $\epsilon_F>0$ and small $\gamma$, $T_c$ is strongly suppressed relative to the rest of the phase diagram. This originates from the absence of angular dependence in the bare polarization bubble, analogous to the parabolic case in Eq.~\eqref{eq:Pi2}. For larger $\gamma$, specifically $\gamma/\epsilon_0 \sim 2$--$3$, $T_c$ increases sharply. This enhancement follows from a rapid change in $\Delta V$, which remains nearly zero (i.e., the effective interaction is nearly angle-independent) at small $\gamma$ and large $\epsilon_F$.

\section{Summary and Outlook \label{Sec:Summary}}

In this work, we analyzed the robustness of the Kohn–Luttinger superconducting mechanism in systems with broken spatial symmetries, focusing on materials with Ising and Rashba spin–orbit coupling. Using analytical modeling and perturbative methods, we showed that symmetry breaking does not eliminate the KL instability. The superconducting $T_c$ can both increase or decrease as a function of the symmetry breaking field, which   is affected by multiple important factors, including the pairing interaction, the density of states and  mixing between attractive and repulsive pairing channels. 

This result is confirmed by explicit models. For Ising SOC, rigid spin-split Fermi-surface shifts leave $T_c$ unchanged, while higher-order anisotropies generally enhance $T_c$ at intermediate scales of the symmetry breaking field and only suppress $T_c$ at very large values. The overall behavior remains the same when adding an out-of-plane Rashba coupling. Thus, the absence of spin conservation (and the existence of non-zero quantum geometry) do not fundamentally modify the results.

Overall, superconductivity driven by repulsive interactions remains remarkably robust: attractive pairing channels still exist even when spin and momentum are strongly entangled. Only extreme  (exceeding Fermi energy) symmetry breaking can significantly weaken the instability. This suggests that KL-type superconductivity may occur in a broader class of spin–orbit-coupled materials and heterostructures than previously anticipated, motivating future work on realistic band structures, finite-momentum pairing, and dynamical interaction effects.

 It is interesting to compare the KL-mechanism  with another setting in which superconductivity arises from a purely repulsive interaction: the Anderson–Morel (AM) picture. There, phonon retardation generates a strong frequency dependence of the effective electron–electron interaction, and Eliashberg theory permits a superconducting solution even when the interaction is repulsive at all frequencies. As in the KL mechanism, the gap function acquires a sign change, in this case as a function of frequency, but such a solution exists only when the Migdal parameter, the ratio of the Fermi energy to the Debye frequency, is sufficiently large ~\cite{Anderson-Morel,marsiglio2020eliashberg,pimenov2022twists}.

 The key distinction of the AM from the KL mechanism is in the origin of the effective attraction and the logarithmic divergence of the pairing susceptibility. In the KL problem, the interaction varies as a function of the angle along the Fermi surface, while the logarithmic divergence arises from integrating over momenta perpendicular to it. These two ingredients factorize: angular dependence shapes the pairing channel, and the perpendicular momentum integral supplies the universal logarithm responsible for the instability. In the AM case, by contrast, both the effective attraction and the logarithmic enhancement   are due to the same variable – frequency. As a result, strong mixing between high- and low-frequency components can overwhelm the attractive contribution generated by retardation, allowing repulsion to prevent a superconducting instability even when parts of the effective interaction are nominally attractive~\cite{Dalal2023}. This structural difference explains why the KL instability is  more robust against perturbations than its AM cousin.

\section{Acknowledgments}
We thank Avraham Klein and Erez Berg for helpful discussions. JR acknowledges funding by the Simons foundation and the ISF under grant No. 915/24.

\begin{appendix}

\appendix
\section{Diagonalization of the BCS Hamiltonian in the case of Ising spin-orbit coupling}\label{app:gap_eq_Ising_SOC}

To diagonalize the mean-field BCS Hamiltonian~\eqref{Eq:HBCSIsing}, we use the following Bogoliubov transformation: 
\be  
\left\{ \begin{matrix} c_{\bs k \ua} & = & u_{\bs k}b_{\bs k +} + v_{\bs k} b^\dagger_{-\bs k -}, \\ c_{-\bs k \da} & = & u_{\bs k}b_{-\bs k -} - v_{\bs k} b^\dagger_{\bs k +},  \\ c^\dagger_{\bs k \ua} & = & u_{\bs k}b^\dagger_{\bs k +} + v_{\bs k} b_{-\bs k -}, \\ c^\dagger_{-\bs k \da} & = & u_{\bs k}b^\dagger_{-\bs k -} - v_{\bs k} b_{\bs k +},
\end{matrix} \right. 
\ee 
where coefficients $u_{\bs k}$ and $v_{\bs k}$ can be chosen real, for the real choice of $\Delta_{\bs k} = \Delta_{\bs k}^*$. Operators $b_{\bs k \pm}$ and their hermitian conjugates satisfy standard fermionic anti-commutation relations. Upon diagonalization, we find 
\be  
H_{BCS} = E_0 +  \sum_{\bs k \in \text{FS}\uparrow} E_{\bs k} \left(b^\dagger_{\bs k +} b_{\bs k +} + b^\dagger_{-\bs k -} b_{-\bs k -}\right),
\ee 
with $E_0$ is the condensate energy and the quasiparticle spectrum
\be  
E_{\bs k} = \sqrt{\xi^2_{\bs k} + \Delta^2_{\bs k}}.
\ee 
The transformation coefficients are given by
\be  
\left\{ \begin{matrix}\displaystyle u^2_{\bs k} - v^2_{\bs k} & = & \frac{\xi_{\bs k}}{E_{\bs k}} \\[1.em] \displaystyle 2 u_{\bs k} v_{\bs k} & = & \frac{\Delta_{\bs k}}{E_{\bs k}} 
\end{matrix} \right. \quad \Rightarrow   \quad \left\{ \begin{matrix} u_{\bs k} & = & \sqrt{\frac12\left(1 + \frac{\xi_{\bs k}}{E_{\bs k}}  \right) }\\ v_{\bs k} & = & \sqrt{\frac12\left(1 - \frac{\xi_{\bs k}}{E_{\bs k}}  \right) } 
\end{matrix} \right.. 
\ee 
Along with identity $E_{\bs k +} = E_{-\bs k -} = E_{\bs k}$, this transformation converts the self-consistency condition~\eqref{Eq:Deltadef} into the gap equation~\eqref{eq:gapeqtoytoy}.



\end{appendix}
\bibliography{KohnLuttinger}

\end{document}